\documentclass[conference]{IEEEtran}
\IEEEoverridecommandlockouts
\usepackage{cite}
\usepackage{amsmath,amssymb,amsfonts}
\usepackage{algorithmic}
\usepackage{graphicx}
\usepackage{textcomp}
\usepackage{xcolor}
\usepackage{url}
\usepackage{hyperref}
\def\BibTeX{{\rm B\kern-.05em{\sc i\kern-.025em b}\kern-.08em
    T\kern-.1667em\lower.7ex\hbox{E}\kern-.125emX}}

\usepackage[disable]{todonotes}

\renewcommand\fbox{\fcolorbox{lightgray}{white}}

\begin{document}

\title{mosaiks are made of tesserae:\\GUI design for a co-simulation framework
\thanks{
This research has been partly funded by the Deutsche Forschungsgemeinschaft (DFG, German Research Foundation) under the National Research Data Infrastructure – NFDI4Energy – 501865131. 
\\
\copyright2026 IEEE.~https://doi.org/10.1109/OSMSES54027.2022.9769116
Personal use of this material is permitted. Permission from IEEE must be obtained for all other uses, in any current or future media, including reprinting/republishing this material for advertising or promotional purposes, creating new collective works, for resale or redistribution to servers or lists, or reuse of any copyrighted component of this work in other works.}
}

\author{\IEEEauthorblockN{Eike Schulte, Jan Sören Schwarz, Malte Stomberg, \\Sharaf Alsharif, Danila Valko, Jirapa Kamsamsong}
\IEEEauthorblockA{\textit{R\&D Division Energy} \\
\textit{OFFIS Institute}\\
Oldenburg, Germany \\
mosaik@offis.de}
}

\maketitle
\begin{abstract}
In a mosaic, a tessera is a single stone.
We introduce \textit{tesserae} for the co-simulation framework mosaik, where they are sets of entities.
They allow for a visual, intuitive, and yet systematic description of simulation scenarios by allowing their entities to be created together and the entities of two tesserae to be connected simultaneously, while ensuring that multidirectional data-flow between tesserae remains consistent without further manual synchronization.
We further present an extension of mosaik by a graphical user interface (GUI) based on these tesserae, enabling the drag-and-drop creation of co-simulation setups and their execution.
The GUI aims to make mosaik more accessible to users previously excluded by its script-based nature.
At the same time, it preserves mosaik's flexibility, extensibility, and modular architecture.
\end{abstract}


\section{Introduction}
\label{sec:introduction}
Co-simulation is a technique that allows the user to analyze complex systems, such as smart grids, by linking multiple different, domain-specific simulators together.
Each linked simulator takes care of its field of expertise while synchronizing and exchanging data with the others.
Usually, the simulation is orchestrated by one central component.
Our open-source framework \textit{mosaik}~\cite{Steinbrink2019,Ofenloch2022} provides such coordination and data exchange for large-scale multi-energy scenarios.
It is capable of handling scenarios for vertically and horizontally scaled cyber-physical energy systems (CPES), supporting the scaling of components from a few to thousands.

Creating and configuring scenarios using mosaik is currently realized by writing \textit{scenario scripts} in Python which represents a significant entry barrier for many potential users from the energy and engineering domains.
At the same time, programmatic interfaces and script-based scenario development shine when scenario authors must instantiate and manage large numbers of entities, often in the hundreds or even thousands~\cite{Barbierato2022}.
This is common in power-grid modeling, which typically involves many buses, loads, generators, and distributed energy resources.
Scenario authors can use the capabilities of the scripting language to systematically create these entities and to set up connections between them, ensuring correct and consistent bidirectional data-flow\cite{Nguyen2017}.
However, scripts do not inherently offer visualization of the scenario and while automatic generation of visualizations is possible, they are frequently either too coarse or too detailed.
This increases cognitive load and makes errors more likely when doing heterogeneous co-simulation design.

\begin{figure}
    \centering
    \fbox{\includegraphics[width=\linewidth]{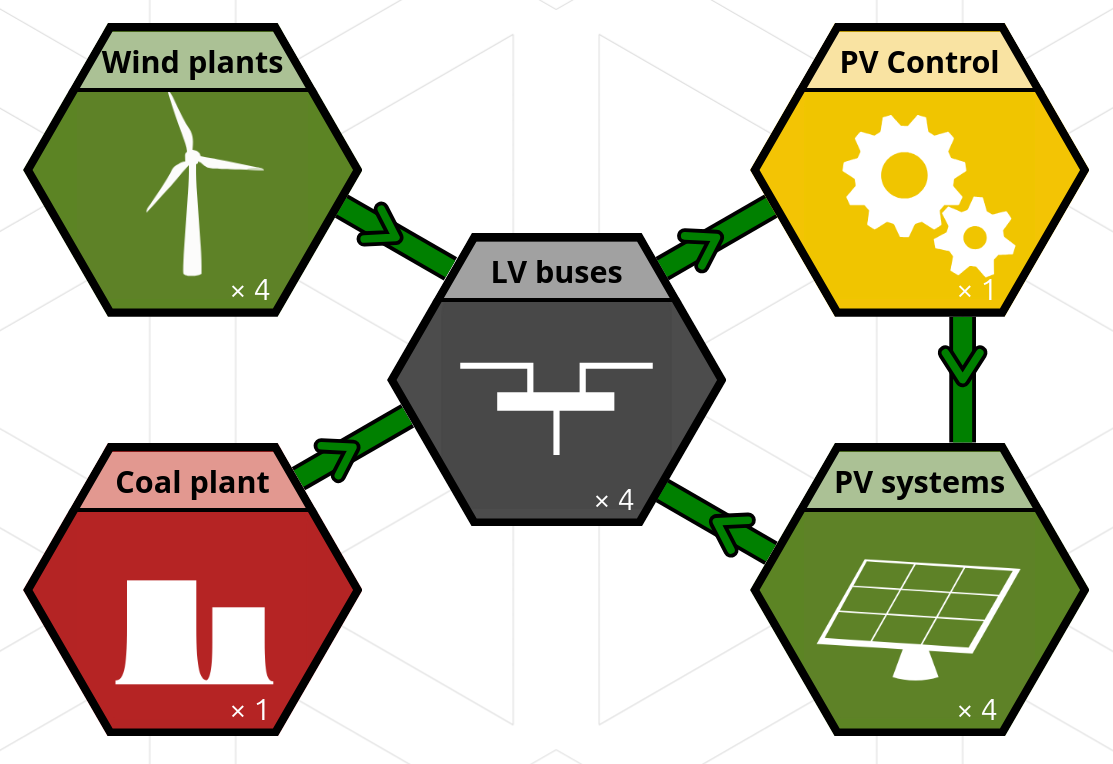}}
    \caption{An example scenario represented using tesserae and connections between them.}
    \label{fig:scenario_overview}
\end{figure}

We propose \textit{tesserae} (singular \textit{tessera}) as a solution to these challenges.
They group fine-grained entities into manageable units, which simplifies both instantiation and connection routines, and enables visualization at an appropriate level of abstraction (see Figure~\ref{fig:scenario_overview}).
Simultaneously, they remain flexible enough to capture all setups commonly implemented in purely script-based workflows.

We further present a GUI for the creation of scenarios in mosaik based on this representation. 
The GUI's design addresses several key use cases: it makes the setup of a mosaik scenario intuitive and easy by allowing the drag-and-drop instantiation of simulators, entities (via tesserae) and the connections between them, and it provides a clear overview of the current scenario setup.
Additionally, it provides the base for more features in the future, like integrating tools for both live-analysis during simulation and efficient analysis of results, and creating an accessible platform that encourages collaborative work among users with diverse backgrounds, e.g., policymakers, grid planners, software developers, and academics.

The implementation of the tessera concept and the GUI are open source and available on GitLab\footnote{https://gitlab.com/mosaik/mosaik-gui}. 

In this paper, we will refer to traditional script-based mosaik as \emph{Python mosaik}, and to the new GUI as \emph{GUI mosaik}.

The remainder of this paper is structured as follows. Section~II reviews related work. Section~III introduces tesserae and scenario descriptions using them in more detail. It also explains how tessera-based scenario descriptions are ``baked'' (i.e., compiled) into Python mosaik. Section~IV details the architectural design of the mosaik GUI. Finally, Section~V concludes the paper and outlines future directions.

\section{State of the art}

There are several co-simulation frameworks and tools in the literature and in practice for analyzing multi-domain energy systems, cyber-physical systems and smart grids. These frameworks differ in several aspects in their orchestrator design, communication paradigm, data exchange methods, and their scalability. An important aspect is also how the co-simulation scenarios are configured and how much effort is required to integrate the simulators' models or entities, which influences the usability of such frameworks\cite{Barbierato2022}.
In the most prominent co-simulation frameworks in the multi-energy domain, the configuration of the scenarios is done through programmatic definitions (e.g., scripting interfaces and JSON/YAML/XML files). For example, in \textit{HELICS}~\cite{HELICS}, the scenario configuration is done with a declarative JSON structure, which reduces the implementation complexity on one hand, but on the other hand is also prone to errors when instantiating and connecting a large number of federates, which is a typical case in the multi-domain energy systems. 
Another example is the use of \textit{AIOMAS}~\cite{scherfke2014aiomas} for co-simulation (as in~\cite{Barbierato2022}). \textit{AIOMAS} is a Python library for multi-agent systems, where the scenario configuration is fully programmatic in this case, and the connectivity is embedded within the agents' definitions. While this provides high flexibility, it significantly increases the configuration effort.

Other frameworks were found to provide some graphical user interface features. For example, \textit{OpSim}~\cite{tobermann2018opsim} provides a graphical interface for its Master Control Program (MCP), which enables monitoring the running co-simulations, connecting to multiple MCP cores, showing component states and the clocks. However, the scenario configuration is still done manually using XML and CSV files.

Another example is the research prototype software \textit{PEGASE}~\cite{vallee2025}, which aims at high-efficiency co-simulation engine and references a two-part graphical interface, one part that supports classical time series plotting features, and the other part gives the possibility to program an elaborated GUI based on the Qt/QML framework. Also, the software could not be accessed.

\textit{DACCOSIM NG}~\cite{Gomez2019} aims at enabling co-simulation based on the FMI standard. The 2019 release provides a GUI for configuring co-simulations using Functional Mockup Units (FMUs), by drag and drop of the FMUs and manual connection of the input and outputs, and the configuration of the parameters of the co-simulation, such as the step size and the iteration limits. While the framework is useful for coupling FMUs, it lacks the required interoperability for models from different domains, possibly written in different languages or following different standards. Furthermore, the manual FMUs creation and input output connections becomes impractical in large-scale multi-energy systems.

In summary, existing co-simulation frameworks for multi-domain energy systems and cyber-physical systems either still rely on manual or programmatic/file-based scenario configuration, or offer GUIs that are domain-specific or not scalable.
This work addresses this gap by developing a GUI extension for mosaik to improve the usability, reduce configuration errors, and minimize the manual configuration overhead in large-scale co-simulation scenarios.

\section{Scenario descriptions using tesserae}

Traditionally, a mosaik user will describe their simulation in a Python script\,---\,called a \emph{scenario} script\,---\,by making calls to the mosaik scenario API.
This involves creating a \emph{world} object; starting simulation components, called \emph{simulators}, within it; and then setting up these simulators and their connections with further API calls.
Finally, the script will have a call to actually run the simulation.
As the simulators are already running during most of this script, scenario authors can dynamically adapt their scenario to information provided by the simulators, which contributes to mosaik's flexibility.

However, a Python script is not well suited for the GUI's goal to provide a clear visual representation that can be edited and reliably saved and opened.
So a more structured representation is required.
At the same time, not too much of mosaik's flexibility should be sacrificed in the process.

The abstract and concrete scenario descriptions in the following subsections are our answer to this.
They turn out to be useful outside of the GUI as well\,---\,for example, to run a GUI-designed scenario on a high-performance machine without a GUI.
Therefore, they are published independently from the GUI as \emph{mosaik-orbit} on the Python Package Index, PyPI.

\subsection{Abstract scenario descriptions}

Each simulator in mosaik provides one or more \emph{models}, representing the types of things it can simulate.
In Python mosaik, the scenario author instantiates these models in their scenario script, creating potentially hundreds of \emph{entities}.
For example, a simulator for renewable energy sources might provide both photovoltaic (PV) and wind turbine (WT) models.
In the simulation, many PV systems and wind turbines might be used, each of which would be an entity of the respective model.
The scenario author sets up the data-flow between simulators by connecting individual entities; usually using some form of looping construct to keep the script manageable.

The levels of detail provided by this are unsuitable for the GUI.
Individual entities are too fine-grained.
Working on this level\,---\,without access to a looping construct\,---\,would require the user to make hundreds of connections by hand, being careful to keep bidirectional data-flow intact:
For example, PV entities might be connected to bus entities in the grid simulator in a two-way fashion and the user would need to make sure that they connect the same pair of entities in both directions.
On the other hand, models are already too coarse:
Again, in our PV example, not every bus in the grid will have a PV system attached, even though all of them will be entities of the model \emph{Bus}.

To address this, the mosaik GUI introduces the concept of \emph{tesserae}.
In a mosaic, a tessera is a single stone.
In the mosaik GUI, a tessera is a specification of how to select or create entities of a specific model in a given simulator.
They are represented as hexagons displaying their name, an icon, and the number of entities they consist of.
Currently, the following two methods of specifying entities are supported:
\begin{itemize}
    \item Create new entities for the tessera (of the specified model in the specified simulator).
    The number of entities can be fixed, or it can be given by the size of another tessera in the simulation.
    \item Select entities created elsewhere in the simulation.
    The entities can be filtered based on their entity IDs and on their so-called ``extra info"\,---\,essentially, data provided by simulators for the express purpose of filtering entities.
\end{itemize}
A tessera can combine multiple instances of these \emph{entity sources} to specify its entities, provided all of them share the same model.

Having created tesserae, \emph{connections} can be established between them.
A connection specifies its source and target tesserae, which attributes to connect (attributes being the individual inputs and outputs of entities), and which \emph{relation} to use for the entities.
Just as there are multiple types of entity sources, there are several types of relations:
\begin{itemize}
    \item The empty relation, not actually connecting anything, serves mostly as a default value.
    \item The one-to-one relation requires both of the connected tesserae to have the same number of entities.
    Each entity is connected to the matched entity in the other tessera.
    \item The random relation connects each entity in the source tessera to a random entity in the target tessera.
    It can be specified whether target entities may repeat.
    \item The many-to-one relation requires the target tessera to consist of exactly one entity.
    All entities of the source are connected to it.
    This is mainly used with a database as the target.
    \item A manual relation can be used to specify the exact pairs of connected entities by hand.
    (Though it is usually better to create tesserae in such a way that one of the other methods can be used.)
\end{itemize}

\begin{figure}
    \centering
     \fbox{\includegraphics[width=\linewidth]{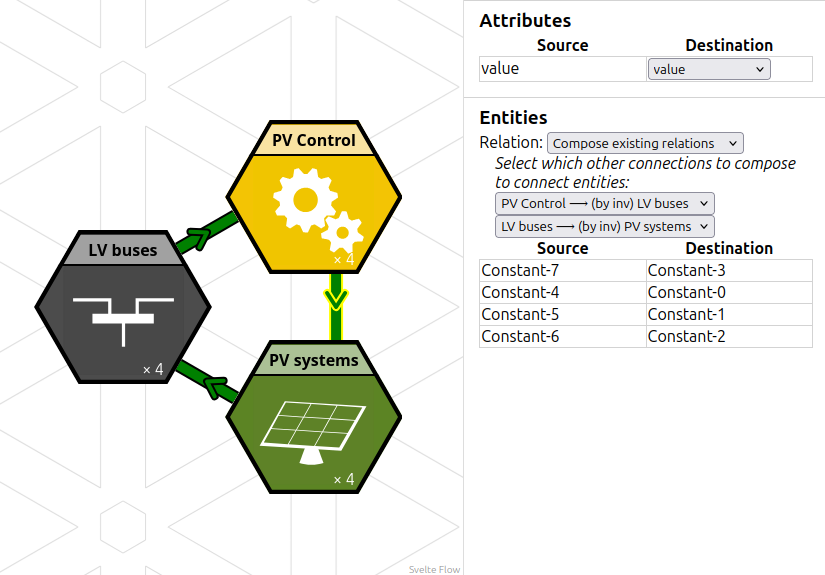}}
    \caption{Triangle of tesserae consisting of PV systems, controllers for those systems, and low-voltage (LV) buses. The settings for the connection from the control systems to the PV systems are shown.}
    \label{fig:example_controller}
\end{figure}

Finally, the relation can also be given as the \emph{composition} of other connections (going both forwards and backwards).
As an example, consider a triangle of tesserae consisting of PV systems, controllers for those systems, and buses (see Figure~\ref{fig:example_controller}).
Each controller receives the state of a bus, uses this data to send control signals to a PV system, and the PV system sends its power output back to the bus.
In this case, it is important that the connections for all three tesserae are consistent.
This can be achieved by setting up the connections from the buses to the controllers and the controllers to the PV systems using methods from the list above, and then specifying that the connection from the PV systems to the buses should be the (inverse of) the composition of those other connections.
This way, when the relation for either of the former connections is changed, the third connections will stay correct automatically.

\begin{figure*}[ht]
    \centering
    \fbox{\includegraphics[width=\linewidth]{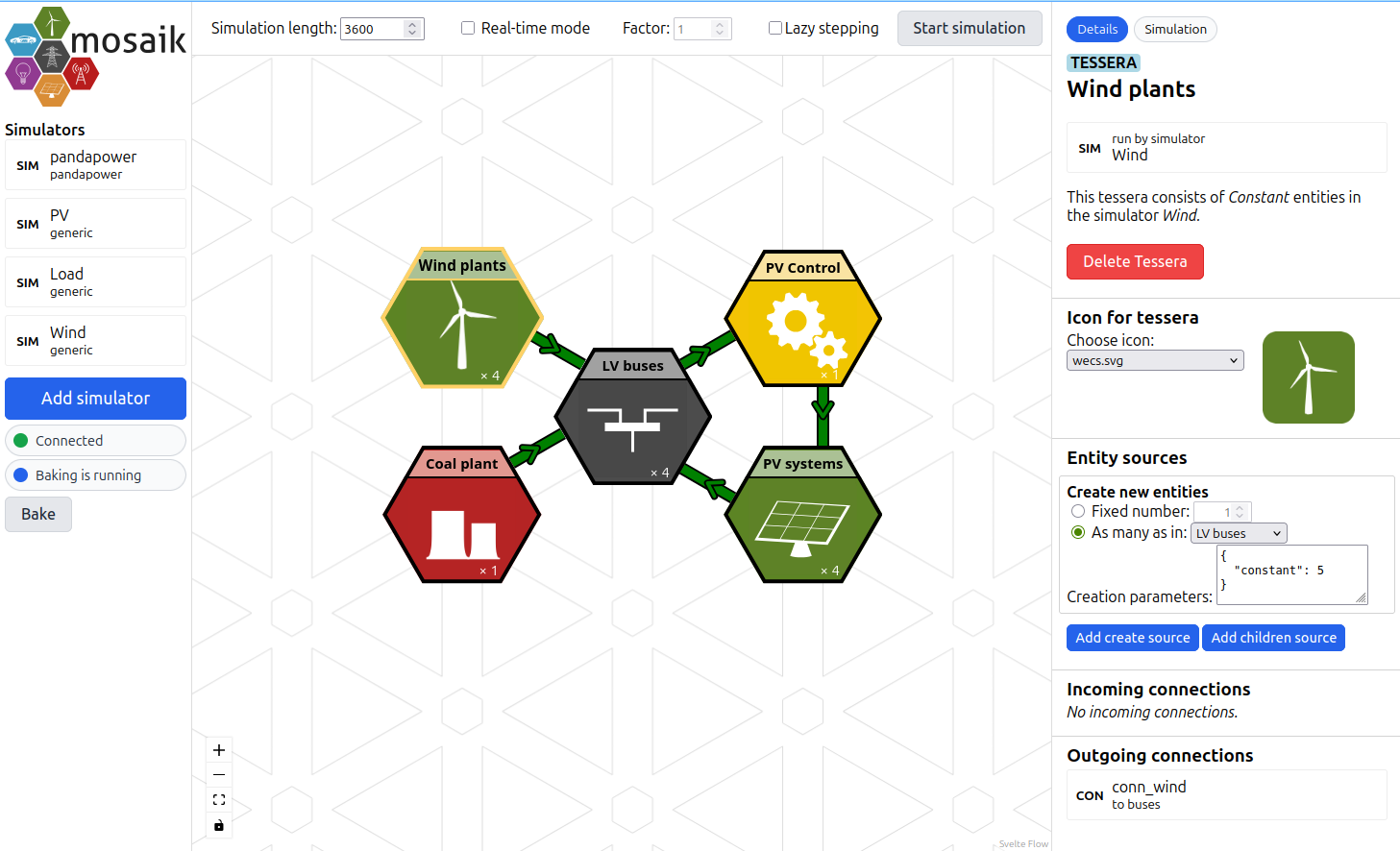}}
    \caption{The example scenario being edited in the mosaik GUI. The wind plants tessera is selected and its settings are visible in the right sidebar.}
    \label{fig:mosaik_gui}
\end{figure*}

\subsection{Concrete scenarios}

The abstract scenario description gets ``baked" into a running instance called an \emph{orbit}.
It consists of the world object from Python mosaik, plus some bookkeeping data to relate it to the scenario description that was used to create it.

During baking, first the actual simulators are started.
Then, the entities are created in them.
This is also when tesserae can determine the actual entities that comprise them.
For example, a tessera might be defined to collect all low-voltage buses in the electrical grid and this can only be done once the grid has been loaded by the grid simulator, as it is the grid simulator that actually loads and parses the file describing the network topology.
Finally, the connections between tesserae are transformed into connections between entities.
Again, it is here that the actual entity pairings can be determined from the abstract relations.

The system is built to be tolerant of errors.
When an element of the simulation fails to initialize, for example due to invalid parameters passed in by the user, this is represented by a ``baking problem'' object.
Only elements that depend on this specific element will then also fail to initialize, while independent elements initialize as normal.
This enables working on partial scenarios in the GUI while still extracting information from the simulators.
For example, due to the existing architecture of mosaik, the models that a generic simulator supports can only be determined once it has been started.
By allowing for partially baked scenarios, we can start a simulator to extract the models, and then allow users to create tesserae for those models.

In the orbit, each of the concrete simulation elements is furthermore associated with the abstract element (and relevant concrete information) used to create it.
When the user changes the scenario, the baking algorithm uses this data to determine whether entities can be reused.
For example, when an entity source specifies that a PV system entity should be created for each low-voltage bus in the grid, and the user then replaces the grid by one with more low-voltage buses, the orbit recognizes that the existing entities can be reused and recreates as many new ones as needed.

The associations are also used to recognize when the user makes changes to the scenario that cannot be replicated on the running world.
For example, mosaik currently has no way of ``unstarting'' a simulator, or deleting entities from it.
(This was simply not a required feature in its pure Python form.)
In these cases, the baking algorithm falls back to a full reset, recreating the entire world from scratch.

Care is taken that the result of baking a given abstract scenario is the same whether starting from a new orbit or from an existing orbit (provided the simulators implement the mosaik API correctly).
In the future, as we extend Python mosaik with new features, like entity deletion, the baking algorithm will be updated to make use of them while preserving this invariant.

\section{Architecture}

\subsection{Overall architecture}

We envision two main use cases for the GUI:
First, it will be integrated into a web-based platform providing ``simulation-as-a-service" (SimaaS).
Second, it should also be usable as ordinary desktop application, running simulations on the user's machine.

These requirements compel us to implement the GUI in a web framework, with the choice having fallen on \emph{Svelte}\footnote{https://svelte.dev}.
The visualization of the scenario as a graph of tesserae is enabled via the \emph{Svelte Flow}\footnote{https://svelteflow.dev} component.

On the other hand, mosaik and vast parts of its ecosystem of simulators are implemented in Python.
As not all simulators are written by us, keeping the core Python-based and not forcing any changes to the simulators were also strict requirements.
Therefore, the mosaik-orbit package is implemented in Python. It provides access to the mosaik orbit and the mosaik world contained within, as well as to the result of the baking via websockets.

\begin{figure}
    \centering
    \fbox{\includegraphics[width=\linewidth]{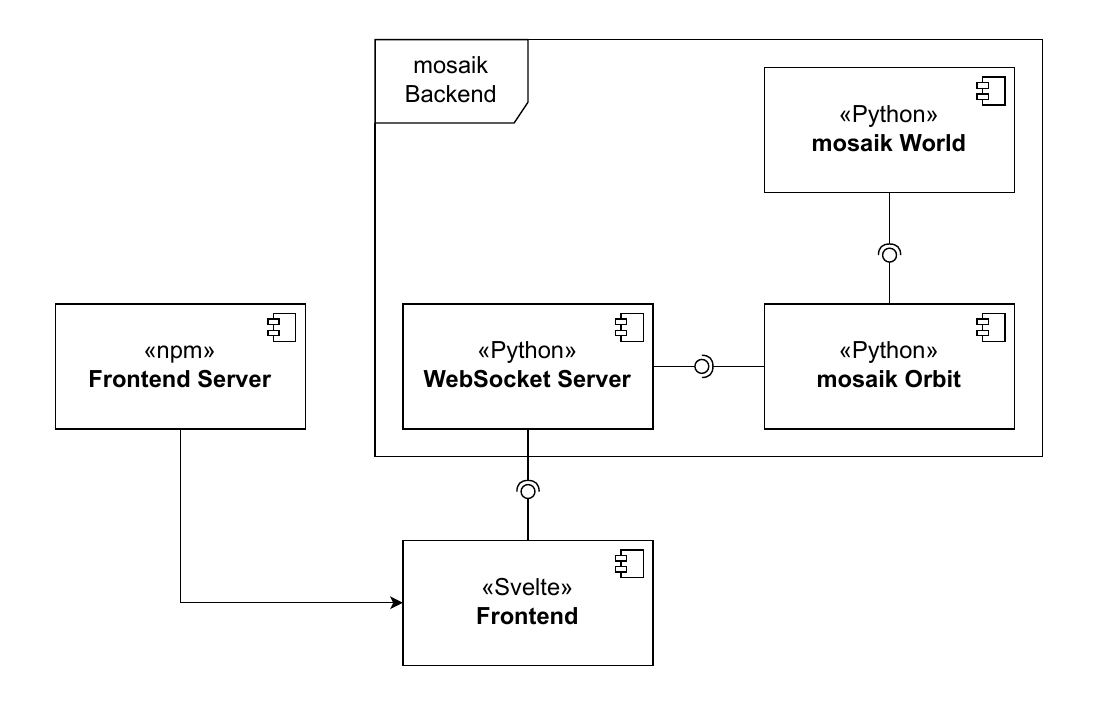}}
    \caption{Architecture of the mosaik GUI.}
    \label{fig:architecture}
\end{figure}

This architecture is visualized in Figure~\ref{fig:architecture}.

The orbit is designed to be lightweight to avoid dependency conflicts with other simulation components.
This way, the entire simulation, including the orbit, can be packaged into a ``mosaik Backend`` container.

In a SimaaS context, these backend containers run on the infrastructure while the frontend runs in the user's browser.
When used as a desktop application, the core is run locally and the frontend is presented using a framework like \emph{Electron}\footnote{https://www.electronjs.org}, instead.
In either case, the frontend is served by a separate frontend server (running on the SimaaS infrastructure or as port of the packaged desktop application).
This part of the system is also responsible for replicating the backend container and for providing the actual frontend with the connection data for it.
(In the current development version, the frontend and backend are instead started manually.)

This architecture allows running multiple separate scenarios, in either use case. Nothing more than creating another instance of the backend container and connecting to it from the frontend is required.

The parts of the user interface that allow saving, loading, and navigating between scenarios must be adapted to each use case.

\subsection{Simulator considerations}

The GUI is designed to capture the usual usage patterns of mosaik.
Its concept of tesserae and connections is applicable to any simulator implementing the mosaik ``simulator API''\footnote{https://mosaik.readthedocs.io/en/latest/mosaik-api/index.html}, and any call to a simulator that can be made in Python mosaik can also be made via the GUI.
In this sense, the GUI provides full access to mosaik.

At the same time, many mosaik simulators will profit from being integrated more tightly into the GUI.
These are the two most relevant ways in which this can be done:
\begin{itemize}
    \item Simulators can provide more structured interfaces to set up both themselves and their entities, e.\,g., by providing dropdown menus for restricted parameters, documenting expected values and their physical units, etc.
    \item Simulators can visualize their state.
    For example, a grid simulator could show a graphical representation of the simulated grid, potentially even updated live with congestion and voltage band violation markings.
\end{itemize}
To support this, we will provide an optional plugin infrastructure to simulator authors.

Which simulators are actually available for scenario authors depends on the simulator Python packages and other software installed in the backend container.
The desktop GUI will allow users to customize those containers to adapt them to their specific needs (in addition to providing a basic container as a baseline).
In the SimaaS context, the available packages must be controlled by the service provider to prevent abuse of the infrastructure.

\subsection{GUI overview}

The mosaik GUI has four main parts, which are shown in Figure~\ref{fig:mosaik_gui}:
\begin{LaTeXdescription}
    \item[Sidebar] The sidebar on the left side lists available simulators and allows to add simulators to the scenario.
    Also, the current state of the connection to the mosaik orbit and the state of the scenario baking process is displayed.
    \item[Scenario view] In the center, the scenario view contains the tesserae and the connections.
    Each tessera is represented by a hexagonal element.
    Tesserae can be moved and connections between them can be added, changed, or removed.
    \item[Details panel] On the right, a second sidebar is providing more details about the currently highlighted element. 
    For example, when a tessera is highlighted, its name, simulator, icon, entity sources, and the entities are shown, as well as its outgoing and incoming connections.
    Using the buttons at the top, the sidebar can be switched to show the progress bar and event log of a running simulation.
    \item[Control bar] On the top, the control bar allows to adjust the scenario parameter, like the length of the simulation or if it should run in real-time mode, and to start the simulation.
\end{LaTeXdescription}

\section{Conclusion and future work}

The new mosaik GUI is a major step to further improve the ease of use of mosaik and make co-simulation accessible to user groups without extensive coding skills.
We presented the GUI utilizing the new concepts of tesserae, entity sources, and relations, which allow to represent even complex co-simulation scenarios.
The current state of development contains the basic functionality for adding simulators, creating tesserae, connect them, and start the simulation.
We published the code open-source and will also extend it in the future.

Once a sufficiently full-featured version of the GUI exists, it\,---\,and with it, the concept of tesserae\,---\,will undergo a user study to determine its accessibility to non-programming users.

There are also ample extension opportunities, several of which were already mentioned:
We plan to provide a plugin infrastructure for simulators and add extended functionality for debugging and live-analysis during the simulation to better understand the behavior of simulators.
Additionally, we will integrate visualization of simulation results and implement a desktop application later.

The mosaik GUI will also be aligned with the NFDI4Energy consortium working on a registry of simulation models that is planned to be integrated in the mosaik GUI to allow direct import of simulators.
Another plan is to integrate the mosaik GUI into the SimaaS hub of NFDI4Energy, which aims to enable the creation and execution of co-simulation on a given infrastructure~\cite{seiwerth_2025_15065996}.
Here, we also envision features for collaborative work on one scenario, with multiple users editing it simultaneously.

\bibliographystyle{IEEEtran}
\bibliography{literature}

\end{document}